\documentclass[a4paper,english,doublecol, linenumbers]{epl2}
\usepackage[T1]{fontenc}
\usepackage[latin9]{inputenc}
\usepackage[active]{srcltx}
\usepackage{verbatim}
\usepackage{float}
\usepackage{textcomp}
\usepackage{amsmath}
\usepackage{amssymb}
\usepackage{graphicx}
\usepackage{esint}

\usepackage{lineno}
\linenumbers
\modulolinenumbers[200]

\title{Geometric phases of water waves}
\shorttitle{Geometric phases of water waves} 

\author{F. Fedele\inst{1}}
\shortauthor{F. Fedele}

\institute{                    
  \inst{1} School of Civil and Environmental Engineering,
 School of Electrical and Computer Engineering\\
Georgia Institute of Technology - Atlanta 30332, Georgia, USA\\
}

\abstract{
Recently, Banner et al. (2014) highlighted a new fundamental property of open ocean wave groups, the so-called crest slowdown. For linear narrowband waves, this is related to the  geometric and dynamical phase velocities $U_d$ and $U_g$ associated with the parallel transport through the principal fiber bundle of the wave motion with $\mathit{U}(1)$ symmetry.  The theoretical predictions are shown to be in fair agreement with ocean field observations, from which the average crest speed $c=U_d+U_g$ with $c/U_d\approx0.8$ and $U_{g}/U_d\approx-0.2$.}

\makeatother

\usepackage{babel}
\begin{document}
\maketitle

\section{Introduction}

Several studies over the past two decades suggest that the initial
speed of breaking crests of dominant open ocean wave groups, or breaker
speeds, are typically $20\%$ lower than expected from linear wave
theory \cite{RappMelville}. A recent study in \cite{Banner_PRL2014}
explains the reduced breaker speed by means of the crest slowdown,
a new fundamental property of non-breaking ocean waves as they occur
naturally, not as uniform wavetrains, but within evolving groups.
Before the focusing point, the crest of the largest wave in the group
slows down as it advances leaning forward, and it becomes symmetrical
as the maximum height is approached. As the wave decays after focus,
the crest accelerates as it leans backward. The crest slowdown and
the forward/backward leaning are generic features of each crest of
water wave groups. They are associated with the energy convergence
in the neighborhood of the focal region, irrespective of whether the
wave evolves to break or not \cite{JFMFedele2014}. These findings
have been validated by means of ocean field observations obtained
by the state-of-the-art stereo imaging (\cite{FedeleOE} and references
therein). In particular, the observed probability density function
of the crest speed $c$ estimated from all the measured crests peaks
at close to $0.8c_{0}$, with $c_{0}$ denoting the phase speed at
the spectral peak \cite{Banner_PRL2014}. 

In this letter, I will show that the crest slowdown is induced by
the natural dispersion of unsteady wave groups. Drawing from quantum
mechanics and differential geometry, it can be explained in terms
of geometric phases and principal fiber bundles \cite{Berry08031984,Simon,Bhandari,Aharonov_Anandan,Anandan_Nature,Segert}.
The remainder of the letter is organized as follows. First, I will
discuss the crest slowdown of deep-water linear waves and briefly
overview the notion of principal fiber bundle and associated dynamical
and geometric phases. The geometric interpretation of the dynamics
of linear narrowband waves and crest slowdown is then presented. This
is followed by a comparison with experimental results and conclusions.

\section{Crest slowdown of deep water waves}

Consider a generic broadbanded linear (small steepness) group of surface
gravity waves traveling on deep waters \cite{JFMFedele2009}

\begin{equation}
\eta=\frac{h}{\sigma^{2}}\int S(\omega)\cos(k(\omega)x-\omega t)d\omega,\label{etas}
\end{equation}
where $\eta$ is the surface displacements, $S(\omega)$ is the wave
spectrum with variance $\sigma^{2}$, dimensionless spectral bandwdith
$\nu=\Delta\omega/\omega_{0}$, dominant wavenumber $k_{0}$ and frequency
$\omega_{0}$, and $h$ is the maximum crest height. In deep waters,
the dispersion relation is given by $k=\omega^{2}/g,$ with $g$ denoting
gravity acceleration and the group velocity $c_{g}=d\omega/dk=c_{p}/2$
is half the phase velocity $c_{p}=\omega/k$ (e.g. \cite{Mei1989}). 

Note that the wave group attains its maximal crest height at the focal
point ($x=0,t=0$) by a constructive superposition of a large number
of elementary waves, whose amplitudes depend upon the assumed spectrum
$S(\omega)$. The speed $c$ of a crest, where $\partial_{x}\eta=0$,
is given by \cite{Longuet-HigginsRandomsurfaces}
\begin{equation}
c=-\frac{\partial_{xt}\eta}{\partial_{xx}\eta}.\label{c}
\end{equation}
From (\ref{etas}), the minimum $c_{min}$ of $c$ is attained at
focus and it is given by the weighted average of the linear phase
speed $C(\omega)=\omega/k(\omega)=g/\omega$ of Fourier waves as

\begin{equation}
c_{min}=\frac{\int S(\omega)k^{2}C(\omega)d\omega}{\int S(\omega)k^{2}d\omega}.\label{c/c0}
\end{equation}
For spectra with Gaussian shape, near the crest maximum
\begin{equation}
\frac{c}{c_{0}}=\frac{c_{min}}{c_{0}}+q(\omega_{0}t)^{2}+O(t^{4})
\end{equation}
where

\begin{equation}
\frac{c_{min}}{c_{0}}=\frac{1+3\nu^{2}}{1+6\nu^{2}+3\nu^{4}}=1-3\nu^{2}+O(\nu^{4}),
\end{equation}
and $q=\nu^{2}+O(\nu^{4}),$ with $c_{0}=\omega_{0}/k_{0}=g/\omega_{0}$
denoting the phase speed at the spectral peak. Thus, the crest speed
tends to dimish as the focal point is reached and the slowdown increases
as the spectrum becomes broadbanded, or $\nu$ increases. Clearly,
the slowdown appears with the modulation of the wavetrain induced
by linear dispersion and it is of $O(\nu^{2})$ for narrowband spectra.
This analysis reveals that wave dispersion and a small but finite
spectral bandwidth is essential for the slowdown to arise. Nonlinear
effects limit the crest slowdown by reducing wave dispersion \cite{JFMFedele2014}.
Indeed, the linear phase speeds $C(\omega)$ of a Fourier wave in
(\ref{c/c0}) tend to increase with its amplitude $A_{f}(\omega)$,
viz. $C(\omega)\rightarrow C(\omega)(1+(kA_{f})^{2})$. 

In this paper, I will study the crest slowdown of linear narrowband
wave groups at deep waters. Without loosing generality, I will consider
a carrier wave $e^{i(x-t)}$ whose phase velocity $c_{p}=1$. In a
reference frame moving at the group velocity $c_{g}=1/2$, the wave
surface displacements $\eta$ is given by 

\begin{equation}
\eta=B(\xi,t)e^{i(x-t)}+\mathrm{c.c.,}\label{eta}
\end{equation}
where $\xi=x-t/2$, and $B$ is a slowly varying envelope, that is
the associated bandwidth $\nu$ must be small. For weak nonlinearities,
$B$ obeys the Nonlinear Schrodinger (NLS) equation \cite{Zakharov1968,Zakharov1999}
or the more general Zakharov equation (e.g. \cite{JFMFedele2014}).
For small wave amplitudes, nonlinearities can be neglected and $B$
satisfies the linear Schrodinger (LS) equation (e.g. \cite{Mei1989})

\begin{equation}
i\textrm{\ensuremath{\partial}}_{t}B=\frac{\delta\mathcal{H}}{\delta\overline{B}}=\partial_{\xi\xi}B.\label{LS}
\end{equation}
Here, $\delta$ denotes variational differentiation, 

\begin{equation}
\mathcal{H}=\int|\partial_{\xi}B|^{2}\, d\xi\label{H}
\end{equation}
is the Hamiltonian and $\overline{B}$ is the complex conjugate of
$B$. $\mathcal{H}$ is an invariant of motion, and so are the action
and momentum

\begin{equation}
\mathcal{A}=\int|B|^{2}\, d\xi,\qquad\mathcal{K}=\mathrm{Im}\int\overline{B}\partial_{\xi}B\, d\xi.\label{AM}
\end{equation}

In the following, I will briefly overview the notion of principal
fiber bundles and geometric phases. Then, I will study the geometric
structure of the state space of the LS equation (\ref{LS}) for the
special class of Gaussian envelopes. 

\begin{figure}[H]
\centering\includegraphics[width=1\columnwidth]{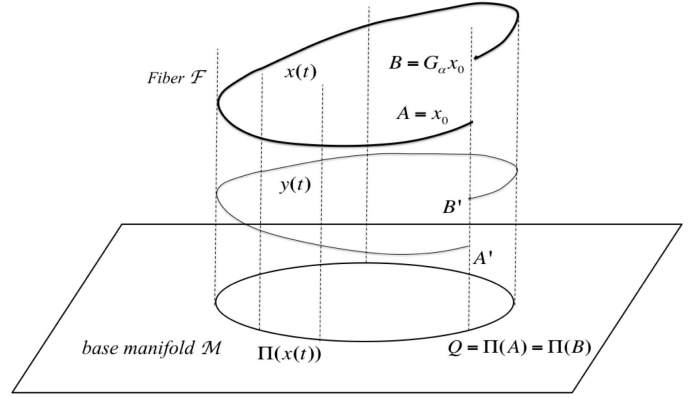} \protect\caption{\label{fig1} Principal fiber bundle: a relative periodic orbit (RPO)
$x(t)$ (AB) reduces to a periodic orbit (PO) in the base manifold
$\mathcal{M}$ by properly shifting the orbit along the fibers $\mathcal{F}$.
The shift is composed by a dynamical and geometric phases. The shift
induced by the dynamical phase yields the orbit $y(t)$ (A'B'), which
is locally transversal to the fibers (parallel transport through the
fiber bundle). A further shift by the geometric phase projects $y(t)$
onto the PO in $\mathcal{M}$. }
\end{figure}

\section{Principal fiber bundles}

Consider a dynamical system with a continuous Lie-group symmetry $G_{\alpha}$
and parameter $\alpha\in\mathbb{\mathcal{\mathbb{R}}}$. The geometric
structure of the associated state space $\mathcal{P}\in\mathbb{R}^{N}$
is that of a principal fiber bundle: a base manifold $\mathcal{M}$
of dimension $N-1$ (quotient space) and one dimensional (1-D) fibers
$\mathcal{F}$ attached to any point $z$ of $\mathcal{M}$. The fiber
$\mathcal{F}$ is the 1-D Lie group orbit $G_{\alpha}z.$ Fibrations
are described by the quadruplet $(\mathcal{P},\mathcal{M},G_{\alpha},\Pi)$
where the map $\Pi:\mathcal{P}\overset{}{\rightarrow}\mathcal{M}$
projects an element $z$ of the state space $\mathcal{P}$ and all
the elements of the group orbit $G_{\alpha}(z)$ into the same point
$\Pi(z)$ of the base manifold $\mathcal{M}$, viz. $\Pi(z)=\Pi(G_{\alpha}z)$.
In $\mathcal{P}$, an orbit $x(t)$ can be observed in a special comoving
frame, from which the motion is an horizontal transport through the
fiber bundle, that is the comoving orbit $y(t)$ is locally transversal
to the fibers (see Fig. \ref{fig1}). The proper shift along the fibers
to bring the motion in the comoving frame is called dynamical phase.
This increases with the time spent by the orbit to wander around $\mathcal{P}$
and system\textquoteright s answer to: \textquotedblright{} how long
did your trip take? \textquotedblright . For example, the translational
shift induced by the constant speed of traveling wave (TW), or relative
fixed point, is the dynamical phase. A TW projects onto the base manifold
$\mathcal{M}$ reducing to a fixed point, whereas a relative periodic
orbit (RPO) reduces to a periodic orbit (PO) (see Fig. \ref{fig1}).
In this case, the shift along the fibers to project a RPO onto the
base manifold $\mathcal{M}$ includes also a geometric phase \cite{Pancharatnam,Simon,Aharonov_Anandan}.
This phase is independent of time and it is induced by the orbital
motion within $\mathcal{M}$, and system\textquoteright s answer to:\textquotedblright which
states have you been through?\textquotedblright . Note that non-periodic
orbits in $\mathcal{M}$ induce also a geometric phase \cite{Anandan_Nature},
that is the motion does not have to be periodic to induce a geometric
drift.

In simple words, geometric phases arise due to anholonomy, that is
global change without local change. The classical example is the parallel
transport of a vector on a sphere. The change in the vector direction
is equal to the solid angle of the closed path spanned by the vector
and it can be described by Hannay's angles \cite{Hannay}. In fluid
mechanics, geometric phases explain self-propulsion at low Reynolds
numbers \cite{Shapere1}. The rotation of Foucault\textquoteright s
pendulum can also be explained by such anholonomy (e.g. \cite{vonBergmann}).
Pancharatnan discovered non-unitary geometric phases for polarized
light \cite{Pancharatnam} as related to a two level quantum system,
and later on Berry rediscovered it for quantum-mechanical systems
\cite{Berry08031984}. Further, unitary geometric phases for waves
have been observed with light \cite{Tomita1986} and elastic waves
\cite{Boulanger}, and these situations are similar to a spin in a
magnetic field \cite{Zwanziger}. 

In the following, I will study the fiber bundle structure of the state
space of the LS model (\ref{LS}) in order to explain the observed
crest slowdown in terms of geometric phases of the orbits in the bundle.

\section{Geometric phases of linear narrowband waves }

The LS equation admits the $U(1)$ group symmetry $G_{\theta}$ and
the translation symmetry $G_{s}$. If $B$ is a solution of (\ref{LS})
so are $G_{\theta}(B)=B\exp(i\theta)$ and $G_{s}(B)=B(\xi+s,T),$
with $\theta$ and $s$ any real number. These symmetries reflect
the invariance of $\mathcal{A}$ and $\mathcal{K}$, respectively
{[}see Eq. (\ref{AM}){]}. $G_{s}$ is not relevant to the crest slowdown,
since it does not induce any dynamical or geometric phase. Hereafter,
I will consider only $G_{\theta}$ and the tangent space to the group
orbit $G_{\theta}B$ at $B$ is 

\begin{equation}
T(B)=(G_{\theta}^{-1}\partial_{\theta}G)B=iB.\label{T}
\end{equation}
To study the qualitative dynamics of a realistic non-breaking ocean
wave group, I consider a finite state space $\mathcal{P\in\mathbb{R^{\mathit{\mathrm{4}}}}}$
of the LS model (\ref{LS}) given by the special class of Gaussian
envelopes

\begin{equation}
B=a(t)\exp(i\xi^{2}\beta(t)-\xi^{2}\gamma(t)),\label{B}
\end{equation}
and respective wave surface displacements

\begin{equation}
\eta=B\exp(i\xi+i\phi(t))+\mathrm{c.c.,}\label{etag}
\end{equation}
where the phase $\phi(t)=-t/2$. From (\ref{LS}), the triplet $z(t)=$$\left[a(t),\;\beta(t),\;\gamma(t)\right]$
satisfies the system of ordinary differential equations (ODEs)

\begin{equation}
\frac{dz}{dt}=N(z),\label{ODEz}
\end{equation}
with $N(z)=\left[2ia(\gamma-i\beta),\:4(\beta^{2}-\gamma^{2}),\:8\gamma\beta\right].$
Here, $a\in\mathbb{C}$, $(\gamma,\beta)\in\mathbb{R}$, and

\begin{equation}
\begin{array}{c}
\mathcal{A}=\sqrt{\frac{\pi}{2}}\frac{\left|a\right|^{2}}{\sqrt{\gamma}},\quad\mathcal{H}=\sqrt{\frac{\pi}{2}}\frac{\left|a\right|^{2}}{\sqrt{\gamma}}\left(\frac{\beta^{2}+\gamma^{2}}{\gamma}\right),\quad\mathcal{K}=0\end{array}\label{AHz}
\end{equation}
are the three invariants inherited from the LS model, where $\left|a\right|$
denotes the absolute value of $a$. The analytical solution of (\ref{ODEz})
follows as 

\begin{equation}
a=\frac{h}{\sqrt{1-2i\nu^{2}t}},\:\gamma=\frac{1}{2}\frac{\nu^{2}}{1+4\nu^{4}t^{2}},\:\beta=-\frac{\nu^{4}t}{1+4\nu^{4}t^{2}},\label{sol}
\end{equation}
with $h$ denoting the maximum wave crest height attained at focus,
and $\nu$ is the spectral bandwidth. The $G_{\theta}$ symmetry is
also inherited, viz. if $z$ is a solution so is $\widetilde{G}_{\theta}z=(a\mathrm{exp}(i\theta),\beta,\gamma)$.
The respective tangent space to the group orbit $\widetilde{G}_{\theta}z$
at $z$ is 

\begin{equation}
\widetilde{T}(z)=(\widetilde{G}_{\theta}^{-1}\partial_{\theta}\widetilde{G})z=(iz_{1},0,0)=i(a,0,0),\label{Tz}
\end{equation}
where $z_{1}$ is the first entry of $z$. The trajectory $z(t)$
associated with $B$ wanders within the state space $\mathcal{P}\in\mathbb{R^{\mathrm{4}}=C\mathrm{x}R^{\mathrm{2}}}$,
which geometrically is a principal fiber bundle $(\mathcal{P},\mathcal{M},\widetilde{G}_{\theta},\Pi)$.
The corresponding symmetry-reduced or desymmetrized orbit $Z(t)$
within the base manifold $\mathcal{M}\in\mathbb{R^{\mathrm{3}}}$
is given by the map 

\begin{equation}
Z(t)=\Pi(z)=(A(t),\beta(t),\gamma(t)),\label{ZT}
\end{equation}
where $A(t)=\left|a(t)\right|$ is the absolute value of $a$ and
$(\gamma,\beta)$ are the same as in $z$. From (\ref{ODEz}), the
motion in $\mathcal{M}$ is governed by the ODEs

\begin{equation}
\frac{dZ}{dt}=N_{0}(Z),\label{ODEZT}
\end{equation}
where $N_{0}(Z)=\left[2A\beta,\:4(\beta^{2}-\gamma^{2}),\:8\gamma\beta\right]$.
From the invariants (\ref{AHz}), the motion occurs on the manifold

\begin{equation}
\begin{array}{c}
\beta^{2}+\gamma^{2}=rA^{4},\end{array}\label{mani}
\end{equation}
where the constant $r=\pi\mathcal{H}/(2\mathcal{A}^{3})$ (see Fig.
\ref{fig2}). The corresponding desymmetrized or symmetry-reduced
envelope 

\begin{equation}
E(\xi,t)=E(Z)=A(t)\exp(i\xi^{2}\beta(t)-\xi^{2}\gamma(t)),\label{EZ}
\end{equation}
and associated wave surface 

\begin{equation}
\eta_{E}(Z)=E(\xi,t)\exp(i\xi)+\mathrm{c.c.}\label{etag-1}
\end{equation}
Given the symmetry-reduced path $Z(t)$ on $\mathcal{M}$, the original
trajectory $z(t)$ can be determined by properly phase-shifting $Z$
along the fibers, viz. 

\begin{equation}
z=(a,\beta,\gamma)=\widetilde{G}_{\theta}Z=(A\mathrm{exp}(i\theta),\beta,\gamma),\label{zt}
\end{equation}
where $\theta(t)$ is the phase-shift (see Fig. \ref{fig1}). Thus,
from (\ref{B}) the associated envelope $B$ in space $\mathcal{P}$
depends upon the desymmetrized orbit $Z$ as

\begin{equation}
B(Z)=E(Z)\exp(i\theta),\label{BE}
\end{equation}
and from (\ref{etag}) the wave surface 
\begin{equation}
\eta=\eta_{E}(Z)\exp(i(\phi+\theta)).\label{etaE}
\end{equation}

\subsection{Dynamical and geometric phases}

To find the phase shift $\theta$ in Eq. (\ref{etaE}), recall that
$E(Z)$ is locally transversal to the fibers. Thus, $\partial_{t}E$
must be orthogonal to the tangent space $T(E)$ {[}see Eq.(\ref{T}){]},
viz.
\begin{equation}
\overline{T(E)}\partial_{t}E=i\overline{E}\partial_{t}E=0.\label{patr}
\end{equation}
The governing equation for $E$ follows from (\ref{LS}) as 

\begin{equation}
i\partial_{t}E-\frac{d\theta}{dt}E=\partial_{\xi\xi}E.
\end{equation}
To impose (\ref{patr}), multiply both members by $\overline{E}$
as 

\begin{equation}
i\overline{E}\partial_{t}E-\frac{d\theta}{dt}\left|E\right|^{2}=\overline{E}\partial_{\xi\xi}E,
\end{equation}
and choose

\begin{equation}
\frac{d\theta}{dt}=\frac{d\theta_{d}}{dt}+\frac{d\theta_{g}}{dt},\label{theta}
\end{equation}
where 

\begin{equation}
\frac{d\theta_{d}}{dt}=-\Omega_{d}=-\frac{\mathrm{Re}\left[\overline{E}\partial_{\xi\xi}E\right]}{\left|E\right|^{2}}\label{thetad}
\end{equation}

\begin{equation}
\frac{d\theta_{g}}{dt}=-\Omega_{g}=-\frac{\mathrm{Im}\left[\overline{E}\partial_{t}E\right]}{\left|E\right|^{2}},\label{thetag}
\end{equation}
Here, $\theta_{d}$ is the dynamical phase, $\theta_{g}$ is the geometric
phase and $\left(\Omega_{d},\Omega_{g}\right)$ are corrections of
the constant dimensionless carrier wave frequency ($=1$). Note that
the 1-forms in (\ref{thetad}) are invariant under the group action,
and $\theta_{d}$ can also be determined from the envelope $B$. In
particular,

\begin{equation}
\frac{d\theta_{d}}{dt}=-\Omega_{d}=-\frac{\mathrm{Re}\left[\overline{B}\partial_{\xi\xi}B\right]}{\left|B\right|^{2}}=\frac{\beta^{2}+\gamma^{2}}{\gamma}=\mathcal{\frac{H}{A}}.\label{thetad1}
\end{equation}
As a result, $\Omega_{d}$ is constant and the dynamical phase increases
linearly with the time spent by the orbit $z(t)$ to wander around
$\mathcal{P}$, viz.

\begin{equation}
\theta_{d}(t)=\theta_{0}+\mathcal{\frac{H}{A}}t,\label{thetad2}
\end{equation}
where $\theta_{0}$ is an arbitrary constant to fix the gauge of freedom
(e.g. \cite{Shapere1}). If I just account for the dynamical $\theta_{d}$,
the comoving orbit $w(t)=(A\mathrm{exp}(-i\theta_{d}),\beta,\gamma)$
associated with the envelope $B_{c}=G_{-\theta_{d}}(B)=B\mathrm{exp}(-i\theta_{d})$
moves through the fiber bundle locally transversal to the fibers.
This motion is referred to as the horizontal transport through the
fiber bundle and it satisfies $\overline{T(B_{c})}\partial_{t}B_{c}=0$
(see Eq. \ref{patr}). 

The comoving envelope $B_{c}$ still experiences a phase shift, the
geometric $\theta_{g}$ given by (\ref{thetag}). Indeed, the desymmetrized
envelope $E$ associated with the orbit $Z(t)$ within the base manifold
$\mathcal{M}$ is obtained by further shifting $B_{c}$ along the
fibers by $\theta_{g}$, viz. $E=G_{-\theta_{g}}(B_{c})=G_{-\theta_{d}-\theta_{g}}(B)$.
Indeed, from (\ref{thetag}) and (\ref{ODEZT})

\begin{equation}
\frac{d\theta_{g}}{dt}=-\frac{\mathrm{Im}\left[\overline{E}\partial_{t}E\right]}{\left|E\right|^{2}}=-\frac{\mathrm{Im}\left[\overline{E}\partial_{Z}E\,\frac{dZ}{dt}\right]}{\left|E\right|^{2}}=-\frac{1}{4\gamma}\frac{d\beta}{dt}\label{thetag1}
\end{equation}
and the geometric phase $\theta_{g}$ follows from the contour integral

\begin{equation}
\theta_{g}=-\int_{\Gamma}\frac{\mathrm{Im}\left[\overline{E}\partial_{Z}E\, dZ\right]}{\left|E\right|^{2}}=-\int_{\Gamma}\frac{1}{4\gamma}d\beta,\label{thetag2}
\end{equation}
where $\Gamma$ is the path generated by the orbit $Z(t)$ in $\mathcal{M}$
(see Fig. \ref{fig2}). Clearly, the geometric phase is independent
of time.

\subsection{Geometric interpretation of the crest slowdown }

The crest speed $c$ of the Gaussian group (\ref{etaE}), or equivalently,
\begin{equation}
\eta=B\exp(ix-it+i\theta(t))+\mathrm{c.c.,}\label{etag-3}
\end{equation}
follows from Eq. (\ref{c}), correct to $O(\nu^{2})$, as

\begin{equation}
\frac{c}{c_{0}}\approx1-\frac{d(\theta_{d}+\theta_{g})}{dt}=U_{d}+U_{g}=1-\mathcal{\frac{H}{A}}+\frac{1}{4\gamma}\frac{d\beta}{dt},\label{thetag2-1}
\end{equation}
where
\begin{equation}
U_{d}=1+\Omega_{d}=1-\mathcal{\frac{H}{A}},\qquad U_{g}=\Omega_{g}=-\frac{1}{4\gamma}\frac{d\beta}{dt}\label{thetag2-1-1}
\end{equation}
are the dynamical and geometric phase velocities associated with $\eta$.
Note that I included the phase velocity of the carrier wave in $U_{d}$
and $\theta_{0}$ in Eq. (\ref{thetad2}) is chosen so that the total
phase shift $\theta$ is null at $t=0$. Note that $U_{d}$ is constant
in time, and the slowdown is purely induced by the geometric component
$U_{g}$ associated with the dynamics within the base manifold $\mathcal{M}$.
This corresponds to a pattern-changing evolution of the wave surface,
which during the focusing event first leans backward and then forward
as discussed below.

\begin{figure}[t]
\includegraphics[bb=0bp 0bp 650bp 450bp,width=1\columnwidth]{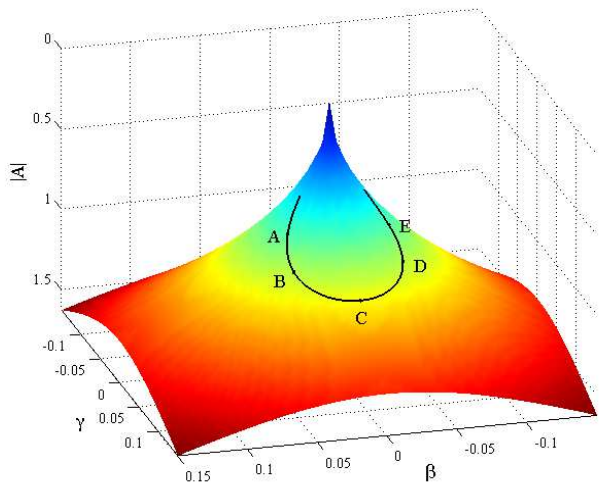}\protect\caption{\label{fig2} Path $\Gamma$of the orbit $Z(t)$ on the base manifold
$\mathcal{M}$.}
\end{figure}

\begin{figure}[!h]
\begin{centering}
\includegraphics[bb=0bp 0bp 700bp 600bp,width=1\columnwidth]{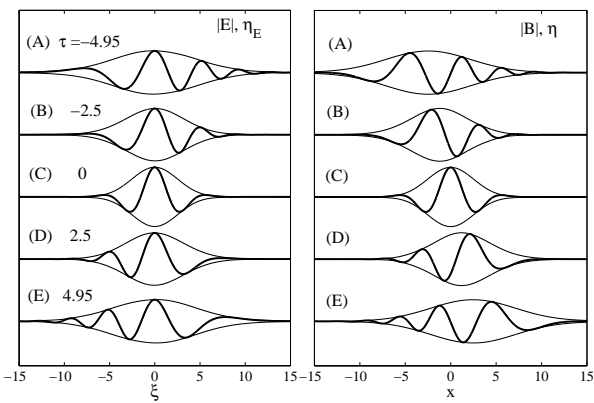}
\par\end{centering}

\protect\caption{\label{fig3}Wave group evolution ($\nu=0.2$): (left) envelope $\left|E\right|$
and wave surface $\eta_{E}$ associated with the path $Z(t)$ on the
base manifold $\mathcal{M}$ at the different time instants or stages
indicated in Fig. \ref{fig2}; (right) same for the envelope $\left|B\right|$
and wave surface $\eta$ associated to the path $z(t)$ on the original
space $\mathcal{P}$. The maximum wave crest is attained at $t=0$
(stage C). }
\end{figure}

The projected wave motion $Z=\Pi(z)$ onto $\mathcal{M}$ is shown
in Fig. \ref{fig2} at different instants of time or stages of the
wave group evolution. The associated desymmetrized envelope $\left|E\right|$
and wave surface $\eta_{E}$ are shown in the left-hand panel of Fig.
\ref{fig3}. The corresponding $\left|B\right|$ and $\eta$ in $\mathcal{P}$
are reported in the right-hand panel of the same figure. The latter
shows that before the focusing point (stage A), the largest wave in
the group advances leaning forward (stage B), and it becomes symmetrical
as the maximum height is approached (stage C). As the wave group grows,
the crest slows down and then it accelerates as it leans backward
(stage D) as the wave decays after focus (stage E). To quantify the
crest profile asymmetries, I define the leaning coefficient $\lambda=d_{R}/d{}_{L}$,
where $d_{R}$ ($d_{L}$) is the distance between the crest location
and the adjacent zero down-crossing (zero up-crossing). Thus, before
(after) focus, $\lambda<1$ ($\lambda$$>1$) because the crest lean
forward (backward) while decelerates (accelerates). At focus, $\lambda=1$
and the crest is symmetric with null acceleration. Note that $\lambda$
is the same for both $\eta$ and $\eta_{E}$. 

\begin{figure}[!h]
\begin{centering}
\includegraphics[bb=-121bp 40bp 700bp 600bp,width=1.05\columnwidth]{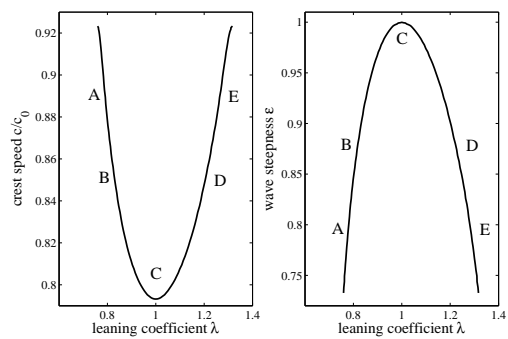}
\par\end{centering}

\protect\caption{\label{fig4} (Left) crest speed $c/c_{0}$ and associated (right)
wave steepness $\varepsilon$ as function of the leaning coefficient
$\lambda$ at the different stages of the wave group evolution (see
Figs. \ref{fig2},\ref{fig3}).}
\end{figure}

In contrast, the dynamics in the base manifold $\mathcal{M}$ does
not present any drift since the desymmetrized wave envelope $\left|E\right|$
does not travel along $\xi$ (see left-hand panel of Fig. \ref{fig3}).
However, the wave surface $\eta_{E}$ features crest asymmetries due
to forward/backward leaning associated with the dynamics within $\mathcal{M}$
(see Fig. \ref{fig2}). This induces the crest slowdown as clearly
seen in Fig. \ref{fig4}, which shows the crest speed $c/c_{0}$ (left-hand
panel) and the wave steepness $\varepsilon=2\pi/Lh$ (right-hand panel)
as function of the leaning coefficient $\lambda$, with $L$ and $h$
denoting the local crest amplitude and wavelength respectively. As
a result, theory predicts that the maximum crest slowdown occurs when
the crest is the largest in the group with maximum steepness and it
has a symmetric profile, viz. $\lambda=1$. In the following, the
theoretical predictions will be compared against ocean field measurements.


\section{Ocean field observations}

The Wave Acquisition Stereo System (WASS) was deployed at the oceanographic
tower Acqua Alta located in the Northern Adriatic Sea, 10 miles off
the coastline of Venice, on 16 meters deep waters \cite{Benetazzo2011}.
Video measurements of the wave surface displacements $\eta$ were
acquired in three experiments carried out during the period 2009-2010
to investigate both space-time and spectral properties of oceanic
waves (\cite{FedeleOE} and references therein). To maximize the common
field of view of the two cameras, WASS cameras were 2.5 m apart, 12.5
m above sea level at 70\textdegree{} depression angle, providing a
trapezoidal area with sides of length 30 m and 100 m, respectively,
and a width of 100 m.

\begin{figure}[!h]
\begin{centering}
\includegraphics[bb=0bp -100bp 700bp 600bp,scale=0.4]{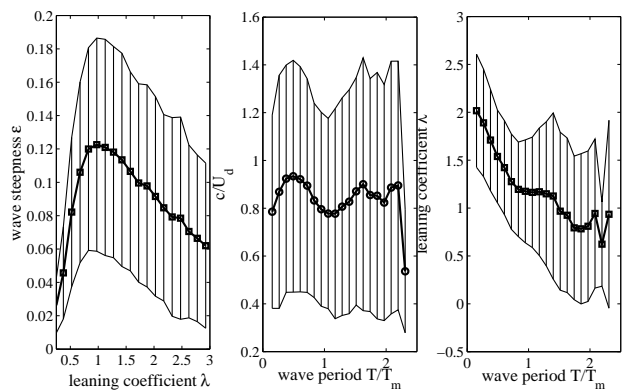}
\par\end{centering}

\protect\caption{\label{fig5} WASS observations: (left) average local steepness $\varepsilon$
versus the leaning coefficient $\lambda$, (center) crest speed $c/U_{d}$
and (right) leaning coefficient $\lambda$ versus the local wave period
$T/T_{m}$, with $T_{m}$ the average wave period. }
\end{figure}

In this work, I elaborated the stereo data acquired during Experiment
2, viz. 21000 snapshots at 10 Hz of the wave surface $\eta(x,y,t)$,
where $x$ is the dominant wave direction and $y$ is that orthogonal
to it. The mean windspeed was 9.6 m/s with a 110 km fetch, and the
unimodal wave spectrum had a significant wave height $H_{s}=1.09\:\mathrm{m}$
and dominant period $T_{p}=4.59\:\mathrm{s}$. Most observed crests
were very steep, with sporadic spilling breaking. The data were filtered
above $1.5$ Hz to remove short riding waves. The speeds $c$ of crests
observed within the imaged area were estimated by tracking the space
crest along the dominant wave direction $x$. Subpixeling reduced
quantization errors in estimating the local crest position and the
leaning coefficient $\lambda$. To estimate the geometric phase velocity
$U_{g}$ of a crest, it is convenient to first estimate the dynamical
phase velocity $U_{d}$ from the observed wave surface $\eta(x,y,t)$,
and then $U_{g}=c-U_{d}$. Note that $\eta$ inherits the translation
symmetry $G_{x_{0}}(\eta)=\eta(x+x_{0},y,t)$ from the symmetries
of the envelope $B$. The dynamical phase velocity $U_{d}$ can be
directly derived by imposing that the material derivative

\begin{equation}
\frac{D\eta}{Dt}=\partial_{t}\eta+U_{d}\partial_{x}\eta
\end{equation}
is the smallest possible in a least-square average, that is 
\begin{equation}
U_{d}(t)=-\frac{\left\langle \partial_{t}\eta\partial_{x}\eta\right\rangle _{x,y}}{\left\langle \left(\partial_{x}\eta\right)^{2}\right\rangle _{x,y}},\label{Uc}
\end{equation}
where the brackets $\left\langle \cdotp\right\rangle {}_{x,y}$ denote
space average in $x$ and $y$. Indeed, Eq. (\ref{Uc}) imposes $\partial_{t}\eta$
orthogonal to the tangent space $(G_{x_{0}}^{-1}\partial_{x_{0}}G)\eta=\partial_{x}\eta$
to the group orbit $G_{x_{0}}(\eta)$ at $\eta$. The left-hand panel
of Fig. \ref{fig5} reports the observed conditional mean value and
stability bands of the wave steepness $\varepsilon$ of dominant local
wave crests as function of $\lambda$, and their amplitude $h>0.3h_{max}$,
with $h_{max}$ denoting the maximum crest height. Note that in average,
the maximum wave steepness is attained for symmetric crest profiles
($\lambda\approx1$). The center panel of the same figure shows the
conditional mean crest speed ratio $c/U_{d}$ as function of the local
wave period $T/T_{m}$, with $T_{m}$ denoting the mean wave period.
From the field observations, $U_{d}=4.22\pm1.09\:\mathrm{m/s}$ and
$c=3.34\pm1.61\:\mathrm{m/s}$. Thus, $U_{d}$ does not vary as much
as $c$ does, as an indication that the variability of $c$ is associated
with that of the geometric component $U_{g}=-0.85\pm1.89\:\mathrm{m/s}$.
The crest speed is the lowest at $T/T_{m}\approx1$ with minimum $c/U_{d}\approx0.8$.
As a result, the geometric phase velocity $U_{g}=c-U_{d}\approx-0.2U_{d}$.
From the right-hand panel of Fig. \ref{fig5}, the average leaning
coefficient $\lambda\approx1.16$ in the range of wave periods $T/T_{m}=0.8-1.2$.
These results are in fair agreement with the above described theory,
which predicts the largest slowdown for symmetric crests ($\lambda=1$).
Clearly, this is an ideal condition as the probability to observe
a symmetric oceanic crest, which is also the maximum of a wave group
(perfect focusing) is practically null (see, for example, \cite{JFMFedele2009}).
Finally, note that the data analysis reveals that the dynamical phase
velocity $\mathit{U}_{d}$ is very close to the reference phase speed
$c_{0}$ at the spectral peak used in \cite{Banner_PRL2014} to find
that the crest speed $c/c_{0}\approx0.8c_{0}$. 

\section{Conclusions}

The crest slowdown is basically a linear phenomenon induced by the
dispersive nature of unsteady wave groups. Drawing from quantum mechanics,
the crest slowdown of linear narrowband waves can be related to the
geometric phase of the orbits in the fiber bundle associated with
the $U(1)$ group symmetry of the wave motion. The role of nonlinear
effects on the crest slowdown is still an open research. In this regard,
initial studies within the framework of the Zakharov equation \cite{JFMFedele2014}
indicate that nonlinearities limit the slowdown effect by dispersion
reduction, that is phase speeds of high-frequency harmonic waves increase
relative to their linear counterparts. Moreover, the crest slowdown
always precedes wave breaking. Further studies are desirable to investigate
how nonlinearities affect the associated dynamical and geometric phases
and their role in the wave breaking process.

\section{Acknowledgments}

WASS experiments at Acqua Alta were supported by Chevron (CASE-EJIP
Joint Industry Project No. 4545093).

\bibliographystyle{eplbib}
\bibliography{geometricphases_citations}

\end{document}